\documentclass[aps,prl,twocolumn,superscriptaddress]{revtex4}
\usepackage{graphicx}
\usepackage{url}
\usepackage{natbib}

\begin{document}

\title{Low Temperature Symmetry of Pyrochlore Oxide Cd$_2$Re$_2$O$_7$}
\author{Jun-Ichi Yamaura and Zenji Hiroi}
\affiliation{Institute for Solid State Physics, The University of Tokyo, Kashiwa, Chiba 277-8581, Japan}
\date{July 8, 2002}
\begin{abstract}
We report the X-ray study for the pyrochlore oxide Cd$_2$Re$_2$O$_7$. Two symmetry-lowering structural transitions were observed at $T_{\rm s1}$=200~K and $T_{\rm s2}$=120~K. The former is of the second order from the ideal cubic pyrochlore structure with space group $Fd\bar{3}m$ to a tetragonally distorted structure with $I\bar{4}m2$, while the latter is of the first order likely to another tetragonal space group $I4_122$. We discuss the feature of the lattice deformation.
\end{abstract}

\maketitle

A series of pyrochlore oxides A$_2$B$_2$O$_7$ show a variety of interesting conductive and magnetic properties, depending on the substitutions of A and B sites~\cite{ref1}. Geometrical frustration on the pyrochlore lattice leads to an exotic ground state such as spin ice for magnetic insulators~\cite{ref2}, while for metallic systems a unique heavy carrier behavior is found in spinel LiV$_2$O$_4$ having a pyrochlore lattice made of V atoms~\cite{ref4}. In the pyrochlore structure, the A site occupied by a rare-earth or a post-transition metal cation forms a distorted (6+2) coordination, while the B site occupied by a transition metal cation forms a distorted octahedron BO$_6$ with an equal B-O distance. The structure comprises two interpenetrating pyrochlore lattices composed of the A or B sites (see Fig.~1).

Recently, Hanawa {\it et al.} and the others reported the first superconductor Cd$_2$Re$_2$O$_7$ ($T_{\rm c}$=1~K) in the family of pyrochlore oxides~\cite{ref5,ref6,ref7}. The compound exhibits two more phase transitions at $T_{\rm s1}$=200~K and $T_{\rm s2}$=120~K, where electrical and magnetic properties change dramatically~\cite{ref8,ref9,ref10}. The former is of the second order, while the latter is of the first order. These transitions are accompanied by structural transitions. Three phases appearing are named as phase I ($T$$>$$T_{\rm s1}$), II ($T_{\rm s1}$$>$$T$$>$$T_{\rm s2}$), and III ($T$$<$$T_{\rm s2}$). Phase I crystallizes in the ideal pyrochlore structure at room temperature with space group $Fd\bar{3}m$. It was reported in the recent structure study that the crystal system is still cubic within the experimental resolution in the wide temperature range down to $T$=10~K~\cite{ref9}. Moreover, taking into account the extinction rule of reflections observed below $T_{\rm s1}$ in the single crystal XRD experiments, a possible cubic space group of $F\bar{4}3m$ was suggested. However, recent Re nuclear quadrupole resonance (NQR) experiments have indicated a lack of a threefold axis below $T_{\rm s1}$ implying that the true symmetry is lower than cubic~\cite{ref11}. In addition, very recent high-resolution XRD measurements using single crystal barely detected a small splitting of cubic Bragg peaks below $T_{\rm s1}$, probably due to tetragonal deformation~\cite{ref12}. In this letter, we report the structural study on the low temperature phases of Cd$_2$Re$_2$O$_7$ by means of X-ray diffraction and discuss possible space groups for them.

\begin{figure}[htb]
\centering
\includegraphics[width=7cm]{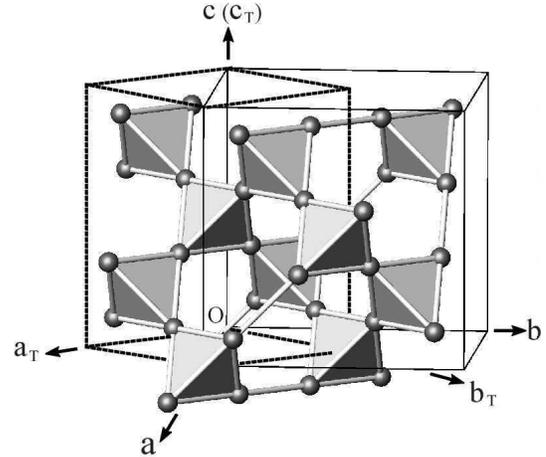}
     \caption{Pyrochlore lattice on Re (=B site) atoms with cubic and tetragonal cell shown. A body-centered tetragonal lattice made by the transformation ${\bf a}_{\rm T}$$=$(${\bf a}-{\bf b}$)/2, ${\bf b}_{\rm T}$$=$(${\bf a}+{\bf b}$)/2 and ${\bf c}_{\rm T}$$=$${\bf c}$ is illustrated by the dashed line. \label{fig:1}}
\end{figure}

A single crystal was synthesized as described in the previous paper~\cite{ref5}. We took oscillation photographs using an imaging plate type Weissenberg camera (Mac Science DIP320V) equipped with a closed-cycle helium refrigerator, which is a specialized system with very low background, in the temperature range of 10~K$-$300~K. The intensity data was collected by a four-circle diffractometer (Mac Science MXC) using a 2$\theta$-$\omega$ scan mode and a CCD area detector (Bruker SMART APEX) equipped with a helium flow type cooler in the temperature range of 85~K$-$300~K. These measurements were performed using a 21~kW rotating-anode X-ray generator with a graphite monochromated Mo-$K$$\alpha$ radiation. The cooling or heating rate was fixed to 0.5~K/min around the first-order transition at $T_{\rm s2}$.

First, we carefully took oscillation photographs in the whole temperature range. Neither additional reflection breaking the face-centered lattice nor peak splitting of reflections were observed. However, the forbidden reflections were found below $T_{\rm s1}$=200~K. Figure 2 (a) shows the temperature dependence of peak intensity for two types of reflections ($0kl:$ $k+l$$\neq$$4n$, $00l:$ $l$$\neq$$4n$), which are extinguished in $Fd\bar{3}m$. They increase gradually below $T_{\rm s1}$ with decreasing temperature, indicating that the transition is of the second order. Moreover, the slight anomaly was observed for some reflections at $T_{\rm s2}$=120~K.

\begin{figure}[tbh]
\centering
\includegraphics[width=7cm]{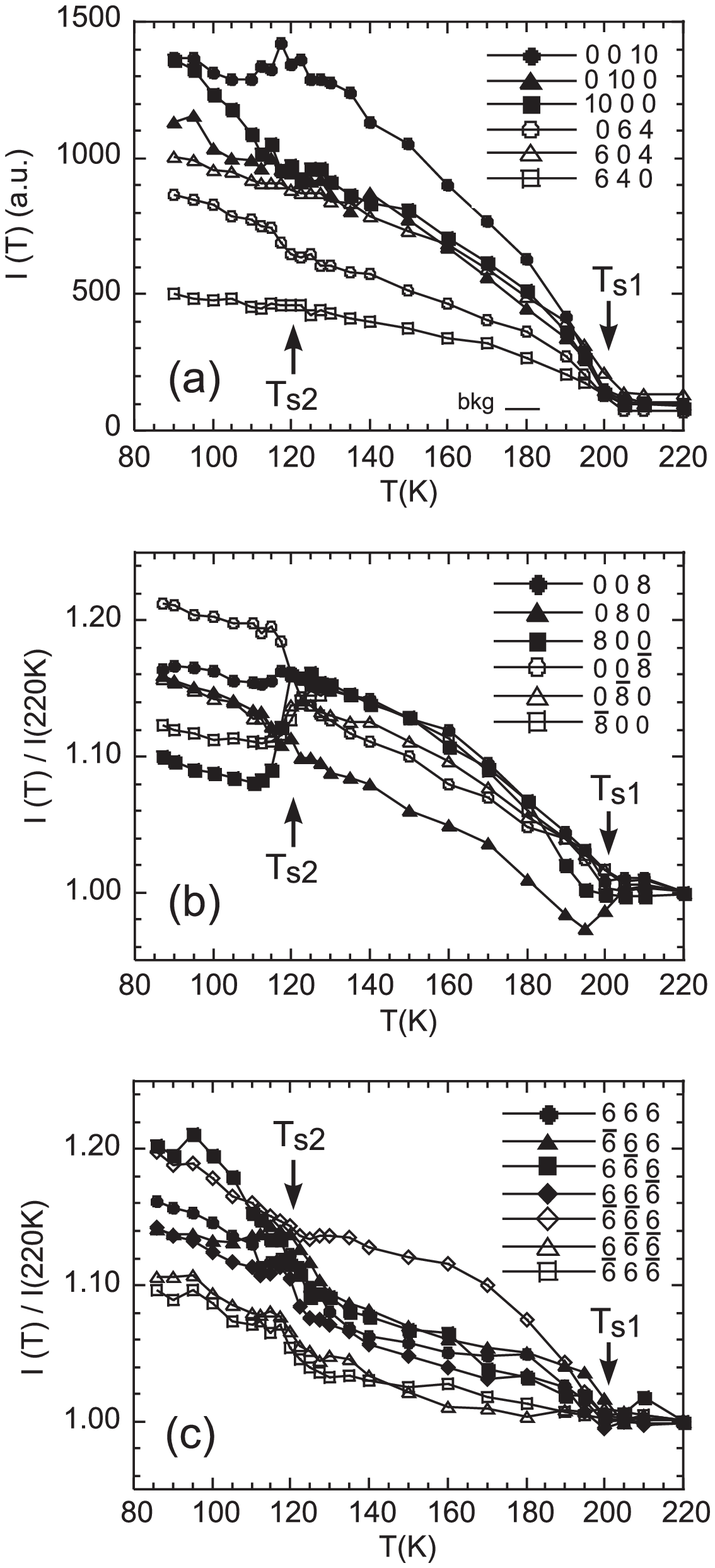}
     \caption{(a) Temperature dependence of peak intensity for reflections ($0kl:$ $k+l$$\neq$$4n$, $00l:$ $l$$\neq$$4n$), which is extinguished in $Fd\bar{3}m$. Temperature dependence of the normalized intensity for 008 (b) 666 (c) with their equivalent reflections in the cubic cell setting. \label{fig:2}}
\end{figure}

Figure 2 (b) and (c) show the temperature dependence of peak intensity for the fundamental reflections 008 and 666 with their equivalent ones, respectively, where the intensity is normalized to the value at 220~K, just above $T_{\rm s1}$, to correct X-ray absorption. It is noted in each case that the equivalent reflections exhibit different intensities below $T_{\rm s1}$ as well as below $T_{\rm s2}$, indicating the reduction of cubic symmetry below $T_{\rm s1}$. Moreover, the sudden step for 008 and their equivalent reflections at $T_{\rm s2}$ suggest that the transition is of the first order. Additionally, the breaking of Friedel's Law was found below $T_{\rm s1}$ as I(008)$\neq$I(00$\bar{8}$) and I(66$\bar{6}$)$\neq$I($\bar{6}\bar{6}$6). This implies the lack of inversion symmetry. Note that Re and Cd atoms have large anomalous scattering factors. The different intensities for the equivalent reflections and the breaking of Friedel's Law were also detected for other many reflections.

Since the transition at $T_{\rm s1}$ is of the second order, we can discuss the symmetry of phase II on the basis of the group-subgroup relations of space group. Figure 3 illustrates the group-subgroup diagram from $Fd\bar{3}m$ to other face-centered cubic or body-centered tetragonal for the maximal translational-equivalent subgroups~\cite{ref13}. Judging from the observed extinction condition, which is compatible with $F\bar{4}3m$ among cubic space groups, and assuming a tetragonal crystal system, one can suggest a possible space group for phase II to be $I\bar{4}m2$ or $I\bar{4}$, that is, the subgroup of $F\bar{4}3m$. Among them, $I\bar{4}m2$ is the most probable space group for phase II, because recent convergent-beam electron diffraction (CBED) study by Tsuda {\it et al.}~\cite{ref14} found the existence of a mirror symmetry at $T$=145~K, which is missing in $I\bar{4}$. Note that, among symmetry operations present for $Fd\bar{3}m$, an inversion symmetry is lost in $F\bar{4}3m$, and additionally a threefold axis is lost in $I\bar{4}m2$. Taking account of no observation of the cubic Bragg peak splitting in our conventional-resolution measurement, the tetragonality of this phase is unusually small, and the resulting deformation of the pyrochlore lattice is minimal. In the $I\bar{4}m2$ model, we estimated the shift of Re atoms from the ideal coordinate (1/4,0,7/8). Assuming that the other atoms are on the ideal positions, the movement of Re atoms is estimated to be $\Delta$$x$$\sim$+0.005 and $\Delta$$z$$\sim$+0.005 (ca. 0.05\AA), using the structure factor at 140~K. Figure 4 shows schematic tetrahedra made of Re atoms and atom positions. The volume of the tetrahedron changes alternatingly without tilting in $I\bar{4}m2$.

\begin{figure}[bth]
\centering
\includegraphics[width=7cm]{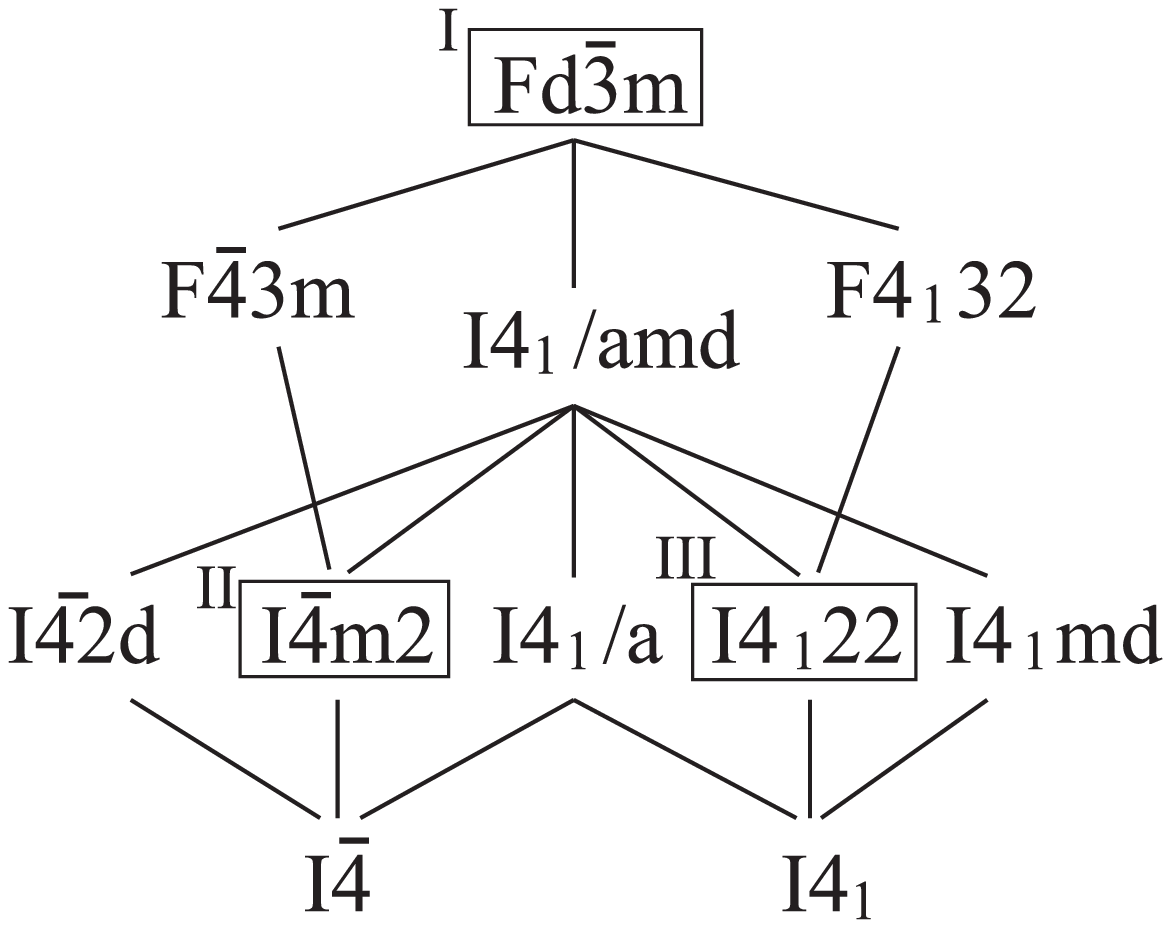}
     \caption{Group-subgroup relations of symmetry reductions for the maximum subgroups, using the conventional space group setting. Symmetry reductions from face-centered to body-centered lattice are described by the translational-equivalent of index 3 ($t3$) with the cell transformation, and the other reductions are described by $t2$. \label{fig:3}}
\end{figure}

\begin{figure}[t]
\centering
\includegraphics[width=7cm]{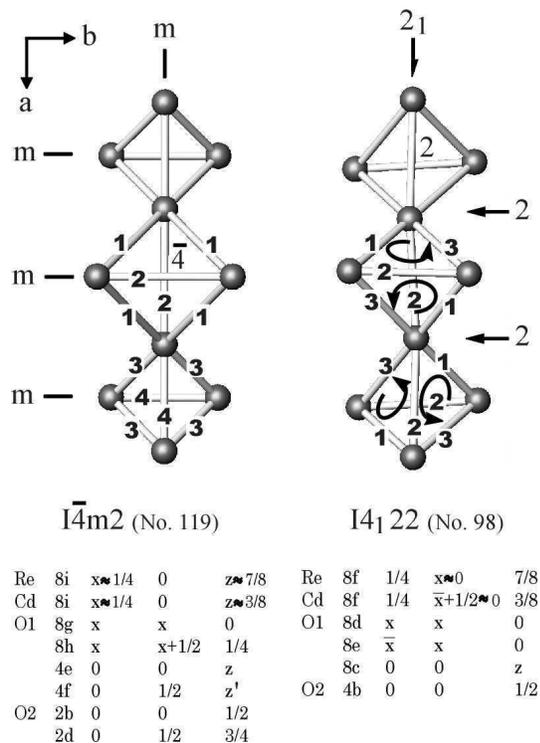}
     \caption{Schematic representations of Re tetrahedra and atom positions for $I\bar{4}m2$ (left) and $I4_122$ (right). Deformation of Re tetrahedra expected for each space group is illustrated exaggeratedly. Numbers on the bonds for $I\bar{4}m2$ and $I4_122$ correspond to four and three kinds of bond lengths, respectively. Rotational arrows imply a possible $bondchirality$ pattern. Parameter $x$,$z$($\approx$$x$$_0$-1/8) and $z$$^\prime$($\approx$$x$$_0$+1/8) on O1 atom is related to the distortion of the ReO$_6$ octahedron. The value of $x$$_0$=0.3186(6) is estimated from our structural data at 300~K, which is different from $x$$_0$=0.3089(7) in Ref. 7. \label{fig:4}}
\end{figure}

Next, we discuss the symmetry of phase III. We checked all the body-centered tetragonal lattices, and found that only the space groups shown in Fig.~3 could construct the pyrochlore structure. Since the transition at $T_{\rm s2}$ is of the first order, the space group at phase III seems not to be included in the subgroup of $I\bar{4}m2$. Therefore, possible space groups are $I4_1/amd$, $I\bar{4}2d$, $I4_1/a$, $I4_122$, $I4_1md$ and $I4_1$. Here, we can exclude $I4_1/amd$, $I\bar{4}2d$, $I4_1/a$ and $I4_1md$, because these are not compatible with the observed extinction condition of reflections with considering a possible crystal twinning. On the other hand, there is no way to distinguish a true one among the other two space groups of $I4_122$ and $I4_1$ from our experiments. However, the CBED experiments suggested that the point group of phase III is 422~\cite{ref14}. This fact makes one space group $I4_122$ selected unambiguously. Thus, it is plausible that a symmetry change from $I\bar{4}m2$ to $I4_122$ occurs at $T_{\rm s2}$, which means missing a mirror symmetry and adding a twofold axis. Such an exchange in symmetry operations is generally seen in a first-order phase transition. It is interesting to note that the Re atom coordinate changes from (x$\approx$1/4,0,z$\approx$7/8) in $I\bar{4}m2$ to (1/4,x$\approx$0,7/8) in $I4_122$. Then, in the $I4_122$ structure, the three kinds of Re-Re bond lengths exist, and the each volume of Re tetrahedon is equivalent. Let's call the three bonds as 1, 2 and 3 in order of bond length. When we trace the bonds of each tetrahedral face as 1$\to$2$\to$3, as shown in Fig.~4, a left-rotation (counter-clockwise) configuration can be defined. Then, all the tetrahedral faces possess the same left-rotation configuration. It is naturally expected that a right-rotation configuration is also realized with equal probability. We call this topological chirality on the Re tetrahedra "$bondchirality$" by analogy with the $R$$-$$S$ convention for a chiral molecule in organic chemistry. The direction of rotation depends on the sign of the Re atom parameter. In a real crystal, two types of domain with the left and right rotations are formed below $T_{\rm s2}$ as a racemic mixture. The Re tetrahedron in $I\bar{4}m2$ is considered to be a disordered state without bondchirality. This means that the transition at $T_{\rm s2}$ is relevant to ordering of the bondchirality.

In summary, we have reported the low temperature X-ray study for the first pyrochlore oxide superconductor Cd$_2$Re$_2$O$_7$. The breaking of the cubic lattice and inversion symmetry are observed below $T_{\rm s1}$=200~K as well as below $T_{\rm s2}$=120~K. From the experimental findings and the group-subgroup relations, it is suggested that the space group changes from $Fd\bar{3}m$ to $I\bar{4}m2$ at $T_{\rm s1}$, and to $I4_122$ at $T_{\rm s2}$. In addition, the characteristic deformations of Re tetrahedron were discussed in each phase.

We are grateful to K. Tsuda, T. Kamiyama, H. Sawa, M. Takigawa, and H. Harima for helpful discussion. This work is supported by a Grant-in-Aid for Scientific Research (C) (No. 12640307) and (A) (No. 407) provided by the Ministry of Education, Science, Sports, Culture and Technology, Japan.

\end{document}